\documentclass[preprint]{myformat}
\usepackage{amsmath}
\usepackage{amssymb}
\usepackage{epsfig}
\usepackage{times}

\def\half{\frac{1}{2}}

\def\0{\varnothing}

\def\al{{\it et. al.~}}
\def\eg{{\it e.g.~}}
\def\ie{{\it i.e.~}}

\title{Phase Transitions in Traffic Models}
\author{E. Levine$^1$, G. Ziv$^1$, L. Gray$^2$, and D. Mukamel$^1$}
\institute{$^1$ Department of Physics of Complex Systems, Weizmann
Institute of Science, Rehovot, Israel 76100. \\
$^2$ School of Mathematics, University of Minnesota, Minneapolis,
Minnesota 55455.} 
\begin{document}

\maketitle

\begin{abstract}
It is suggested that the question of existence of a jamming phase
transition in a broad class of single-lane cellular-automaton
traffic models may be studied using a correspondence to the
asymmetric chipping model. In models where such correspondence is
applicable, jamming phase transition does not take place. Rather,
the system exhibits a smooth crossover between free-flow and
jammed states, as the car density is increased.
\end{abstract}

Traffic flow and the formation of traffic jams have been
extensively studied for many years \cite{Chowdhury00}. A very
useful quantity which has often been used to characterize traffic
flow is the relation between the density of cars in the road and
the traffic throughput. This relation, termed the {\em fundamental
diagram}, was measured empirically in various situations, and was
studied in a large variety of models \cite{Hall, Chowdhury00}.
When the density of cars in the system is low, the traffic flow is
expected to grow linearly with the density. At high densities
traffic jams are formed, lowering down the flow sometimes even to
a complete stop \cite{Brilon, Hall}. This enables one to identify
three regimes in the density-flow plane: a free-flow regime at low
densities; a regime of wide moving jams at high densities; and a
synchronized flow regime, where jams and free-flow coexist, at
intermediate densities \cite{Chowdhury00}. The question whether
the transition from one regime to another is a smooth crossover or
is a result of a genuine phase transition is still not settled in
most traffic models.

In recent years Probabilistic Cellular Automata (CA) models have
been introduced  to analyze traffic flow
\cite{Cremer86,Nagel92,Chowdhury00}. In such models both time and
space are discrete and the physical state of the system (\eg
position and velocity of all cars) is updated simultaneously
according to some update scheme. This provides a rather efficient
way for carrying out numerical studies of the fundamental diagram.
Some traffic CA models were suggested to exhibit phase transitions
\cite{Krauss97,Eisenblatter98,Lubeck98,VDR}. However, the
existence of such a transition can only be explicitly demonstrated
in some limiting cases where certain dynamical processes are
deterministic. The existence of a jamming phase transitions in
more generic cases, where all dynamical processes are
non-deterministic, is still an open question. Such transitions
have been suggested to occur in some models on the basis of
mean-field methods and numerical simulations, which cannot yield a
definitive answer to this question. In this Letter we address this
issue in more detail.

Within a more general framework, the existence of phase
transitions in one-dimensional driven systems has been studied
rather extensively in recent years and several mechanisms for such
transitions have been proposed. Examples include the zero range
process \cite{Evans00,Kafri02A}, two species driven models
\cite{GunterRev} and the chipping model
\cite{Majumdar98,Rajesh02}. The chipping model introduced by
Majundar \al incorporates dynamical processes which are closely
related to those taking place in traffic dynamics, so one would
hope to obtain useful insights for traffic jams from what is known
about the chipping model.

In this Letter we examine the correspondence between traffic CA
models and the chipping model more closely. This correspondence
suggests that for a large class of traffic models with
non-deterministic dynamics, a genuine phase transition is not
expected. Rather, these models exhibit a smooth crossover between
the free flow and the jammed phases. In the following we briefly
review the main known results for the chipping model. We then
introduce a simple CA traffic model for which the correspondence
to the chipping model could be made explicit. Other CA traffic
models which have been introduced and studied in the past are also
examined within this approach.

We start by considering the {\em Chipping Model} (CM). The model
is defined on a periodic lattice, where each site can contain any
number of particles. The dynamics is defined through the rates by
which two nearest neighbor sites containing $k$ and $m$ particles,
respectively, exchange particles:
\begin{align}
\label{eq.CM}
(k,m)\mathop{\longrightarrow}^1(k+m,0)&&(k,m)\mathop{\longrightarrow}^{\omega_L}(k+1,m-1)&&
(k,m)\mathop{\longrightarrow}^{\omega_R}(k-1,m+1)\,.
\end{align}
The first is a diffusion (or coalescence) process\footnote{Here
only a diffusion to the left is considered. One can also introduce
diffusion to the right, without changing the relevant results.}
while the last two processes correspond to right and left chipping
of a particle from one site to the other. It has been shown
\cite{Majumdar98} that if the chipping process is symmetric
($\omega_R=\omega_L$) there is a condensation transition at a
critical density, above which one site becomes macroscopically
occupied. Furthermore, numerical simulations and mean-field
studies show that the probability $P(k)$ of finding $k$ particles
in a site has the asymptotic form $P(k)\sim z^k/k^\tau$ for large
$k$, with $\tau=5/2$. The parameter $z \leq 1$ is determined by
the average particles density and serves as the fugacity. The
condensation transition is a result of the fact that $\tau>2$, for
which the distribution $P(k)$ cannot sustain high densities even
at $z=1$. This transition is analogous to the Bose-Einstein
condensation. In contrast, if the chipping is asymmetric there
exists no phase transition at any density \cite{Rajesh02}. In this
case numerical studies indicate that the domain size distribution
has the same form as above, but here $\tau = 2$. This distribution
remains valid at any density with $z$ approaching~$1$ at high
densities, indicating that no condensation transition takes place.

In the following we argue that the chipping model with an
asymmetric chipping process provides a framework within which a
large class of traffic models can be characterized. Starting from
a particular traffic model we first identify the {\em domains}
which characterize the flow. A domain can either be a low density
segment, termed {\em a gap} or {\em a hole} in some studies ; a
high density segment, termed {\em a jam} ; or a segment of some
other characteristics, defined ad-hoc.
A domain of size $k$ is then associated with a site of
the CM occupied by $k$ particles. One then proceeds by examining
the evolution of the domains, and identifying their dynamical
processes. As will be demonstrated, in many cases these processes
are closely related to the diffusion and the chipping processes of
the asymmetric CM.

\begin{figure}
\twoimages[width=7cm]{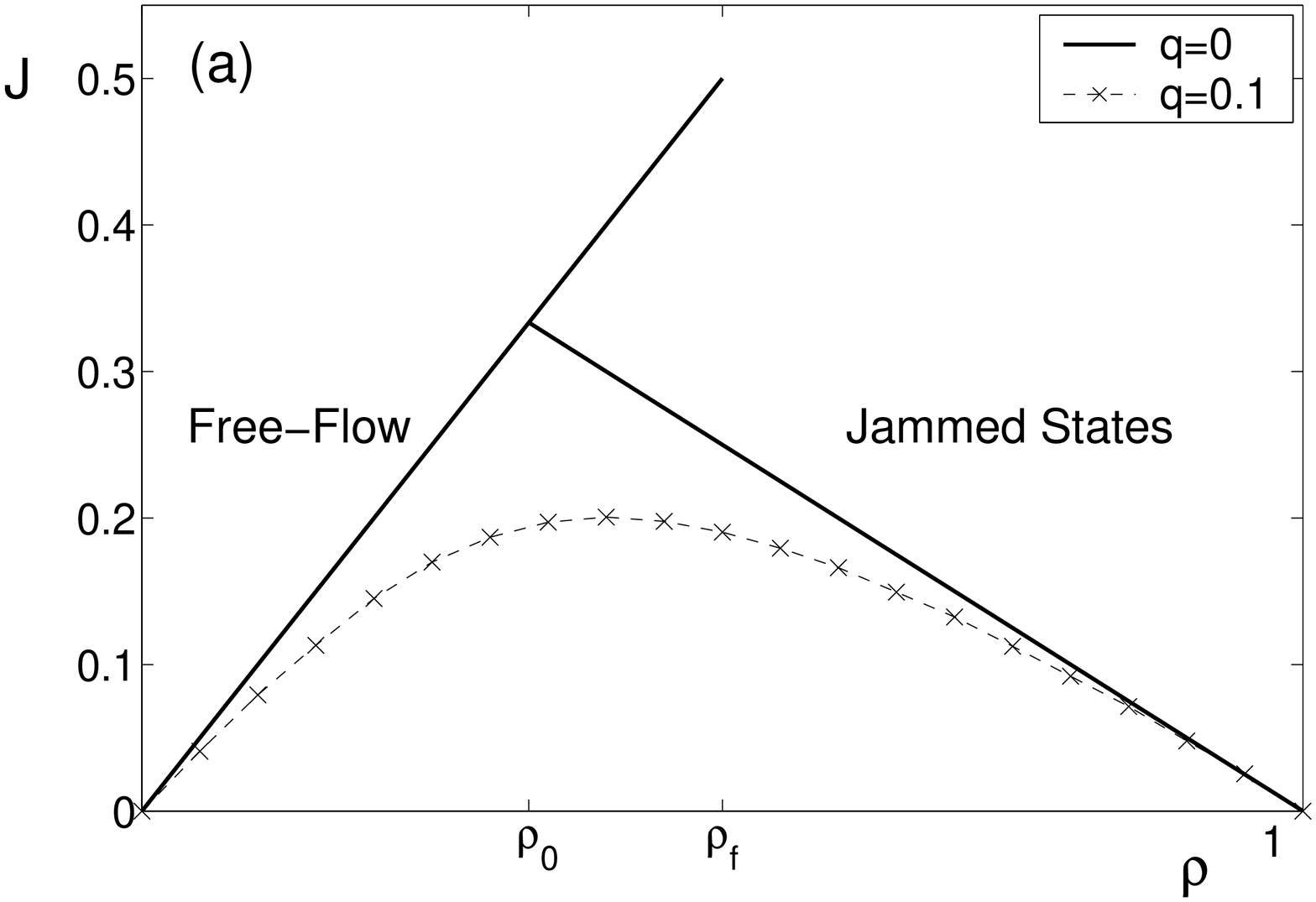}{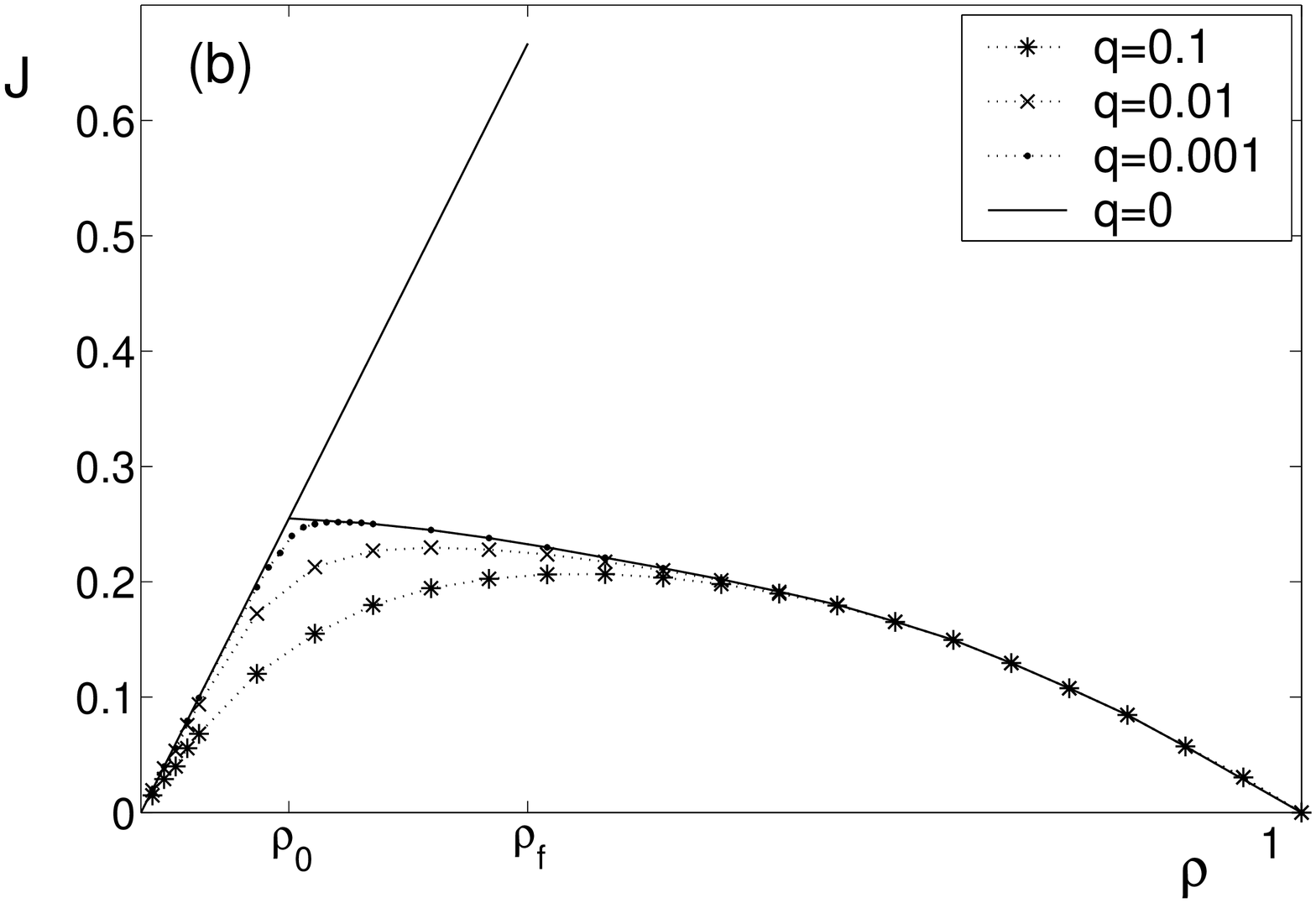}
\caption{Fundamental diagram for the VDB model. Data for $q>0$ was
obtained from numerical simulations of systems of size $500$. (a)
$v_{\max}=1$ and $p=1/2$. (b) $v_{\max}=2$ and $p=0.6$ ; here the
$q=0$ free-flow branch is given by $2\rho$, while the branch of
jammed states in this case was extrapolated from numerical data.}
\label{fig.1}
\end{figure}

We now consider a particular traffic model for which the
correspondence to the CM can be made rather explicit. The model,
referred as Velocity Dependent Braking (VDB), is a variant of the
Nagel-Screckenberg model \cite{Nagel92}. It is defined on a
periodic lattice of size $L$ with $M=\rho L$ cars. Each car is
characterized by a velocity $v_i(t)=0\ldots v_{\max}$ and a
position $x_i(t)$. The dynamics is performed in parallel by first
updating the velocities as
\begin{subequations}
\label{eq.model}
\begin{equation}
\label{eq.modela}
v_i(t+1)=\begin{cases}\min\left\{v_i(t)+1,v_{\max},x_{i+1}(t)-x_i(t)-1\right\}&
\text{with probability } 1-p(v_i(t)) \\
0 & \text{with probability } p(v_i(t)) \end{cases}\,,
\end{equation}
and then increasing the position of each car by its speed,
\begin{equation}
x_i(t+1)=x_i(t)+v_i(t+1)\,.
\end{equation}
\end{subequations}
The braking probability $p(v)$ is defined in terms of the two
parameters of the model, $p$ and $q$, as
\begin{equation}
p(v)=\begin{cases}p&v<v_{\max}\\q&v=v_{\max}\end{cases}\,.
\end{equation}

We proceed by first considering the model in the {\em cruise
control} (CC) limit, $q=0$. In this limit cars residing in dilute
regions of the system move deterministically with the maximal
allowed velocity $v_{\max}$. It is thus straightforward to define
a free-flow domain as a segment which consists of vacancies and
deterministically moving cars. For $\rho\le\rho_f=1/(v_{\max}+1)$
a free-flow steady state exists, where all cars move
deterministically and the current is simply given by
$J(\rho)=v_{\max}\,\rho$. At higher densities such a state does
not exist, local jams are formed and the current is reduced. A
phase transition between the two regimes thus takes place at some
density $\rho_0\le\rho_f$. This particular phase transition is a
result of the fact that due to the deterministic processes, the
free-flow state is an absorbing state which has no dynamics. It is
thus expected to exist only in the CC limit, and to turn into a
smooth crossover for $q>0$, where no absorbing state exists.

It is straightforward to analyze the fundamental diagram in the CC
limit ($q=0$) for the case $v_{\max}=1$.
We find $\rho_f=1/2$ and $\rho_0=(1-p)/(2-p)$. It
can be shown that in the jammed state ($\rho > \rho_0$) the
current is $J(\rho) = \rho_0(1-\rho)/(1-\rho_0)$. For
$\rho_0\le\rho\le\rho_f$ both free-flow states and jammed states
coexist in the thermodynamic limit. In this region a free-flow
state evolves deterministically and jams are never produced.
However, starting from a random initial condition a jammed state
is formed, which slowly evolves towards the free-flow one.
However, the time it takes for a system to reach a free-flow state
increases exponentially with the system size \cite{Ziv04}. Thus in
the thermodynamic limit both the free flow and the jammed phases
exist as stable steady states of the system. For $q>0$ the
$J(\rho)$ curve can be calculated numerically, and it shows no
singularity, as expected. The fundamental diagram of the model for
the case $v_{\max}=1$ is given in Fig.~\ref{fig.1}a.

\begin{figure}
\twoimages[width=7cm]{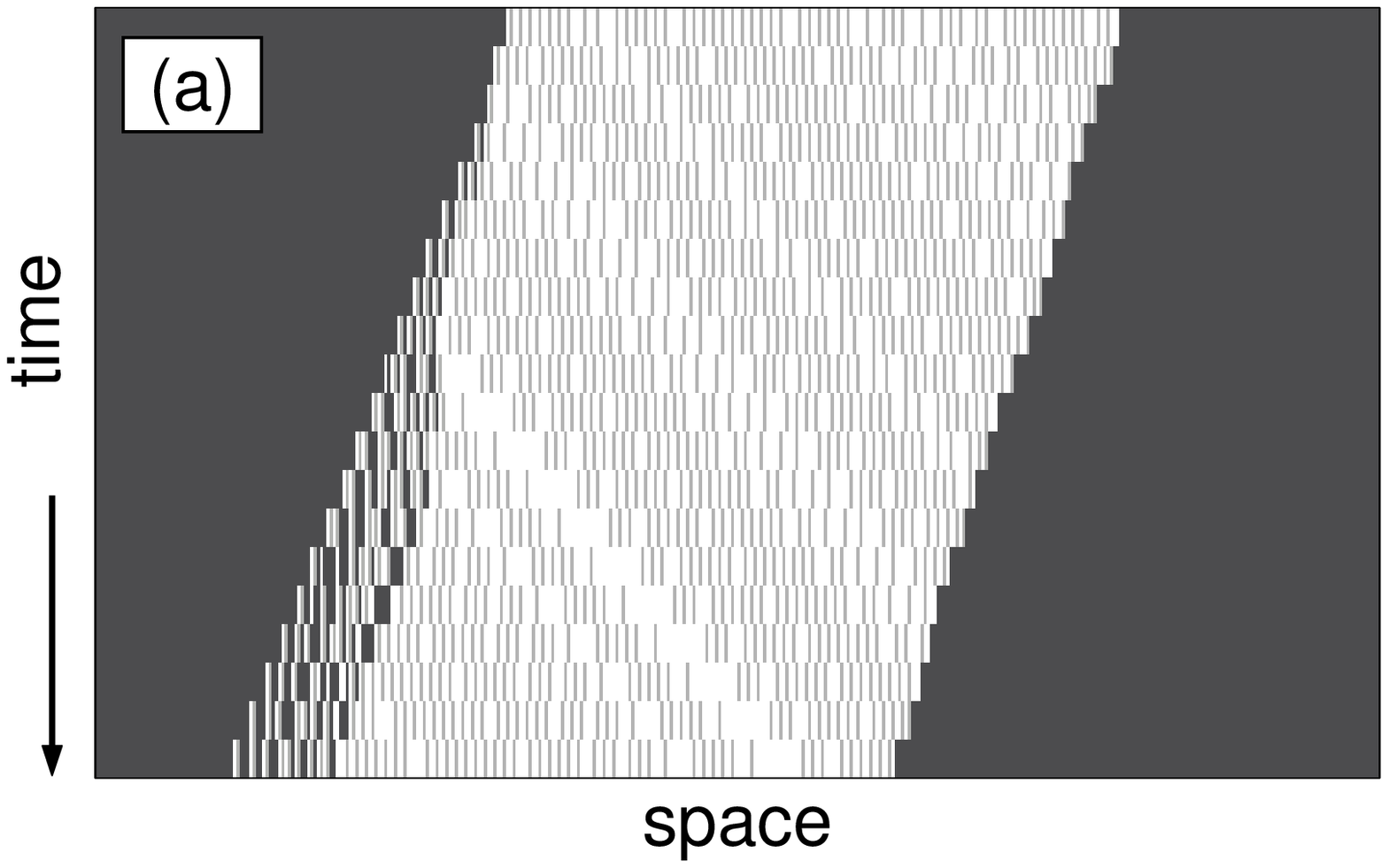}{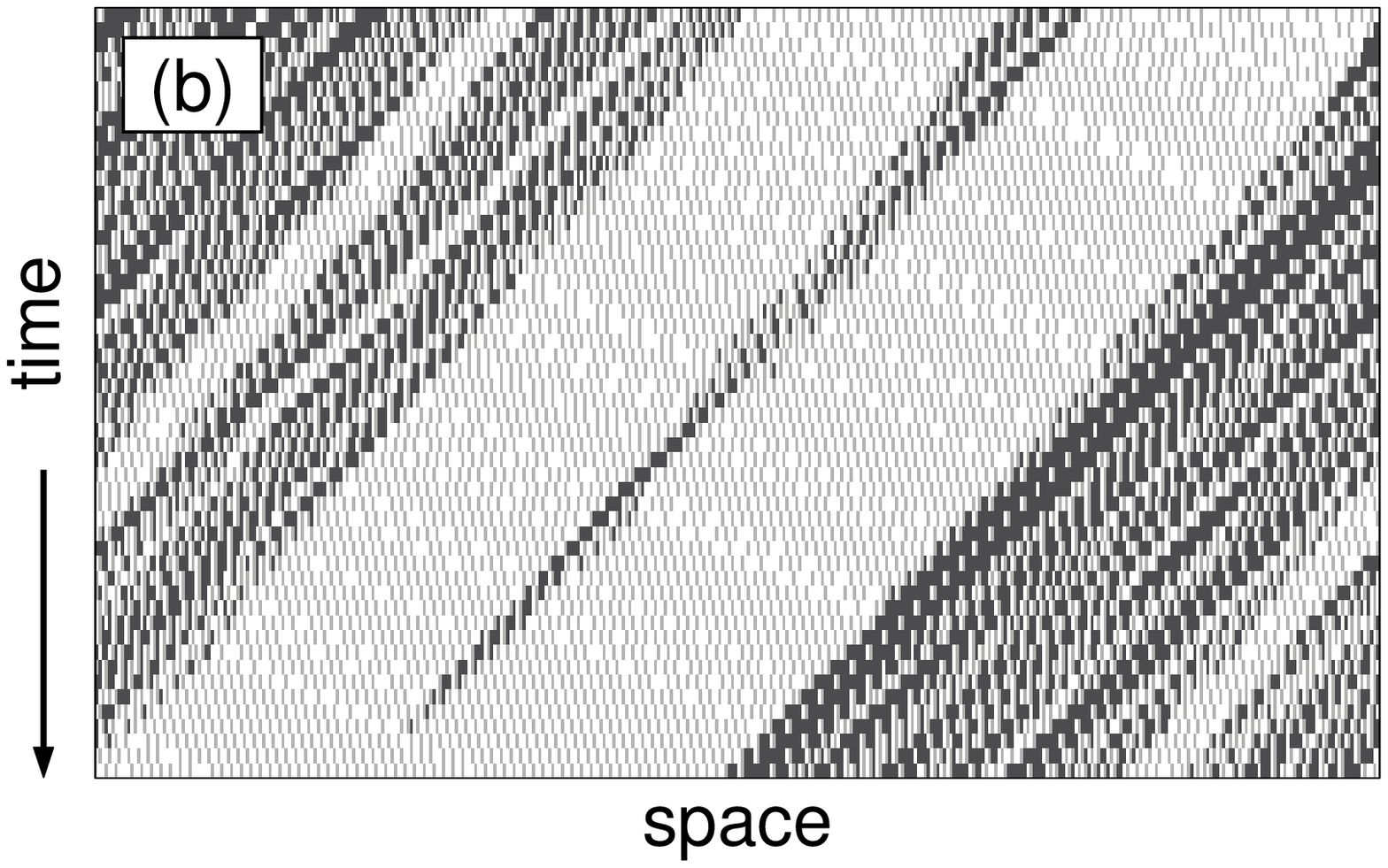} \caption{Space-time
configurations of the VDB model. Stopped cars are in black, moving
cars in gray, vacancies in white. (a) The evolution of a single
domain is characterized by a biased diffusion and chipping of
small domains. Every $5$th sweep is presented. (b) Chipping and
diffusion may lead to coalescence of neighboring domains.
Configurations are presented every 30 sweeps.} \label{fig.config}
\end{figure}

As in many traffic models, the general features of the model are
revealed only at $v_{\max}>1$. We still expect a genuine phase
transition between the free-flow and jammed states in the CC
limit. Again this transition is expected to turn into a crossover
for $q>0$ (see Fig. \ref{fig.1}b). The question is whether there
is another transition at $\rho>\rho_0$ which is not associated
with the existence of absorbing free-flow states, and which may
persist beyond the CC limit, namely for $q>0$.

In what follows we consider $v_{\max}=2$ in the $\rho > \rho_0$
regime. Here no exact solution is available. Instead, we analyze
the evolution of free-flow domains, and examine its correspondence
to the dynamical processes of the chipping model. In Fig.
\ref{fig.config} we present space-time configurations of the model
at $v_{\max}=2$ and $p=0.1$. Focusing on a single domain (Fig.
\ref{fig.config}a) we observe that the domain evolves by biased
diffusion and chipping processes. In Fig. \ref{fig.config}b the
evolution of many domains is depicted, demonstrating exchange of
particles between domains and coalescence of two domains upon
contact.

\begin{figure}
\onefigure[width=14cm]{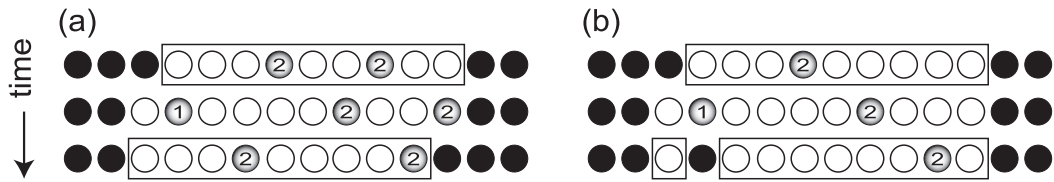} \caption{The $v_{\max}=2$
dynamics of a domain, depicting (a) diffusion and (b) chipping
processes. Stationary cars are marked in black and vacancies in
white. The velocity of moving cars, marked in gray, is indicated.
} \label{fig.cartoon}
\end{figure}

More explicitly, the dynamics of the domains can be linked to the
microscopic processes of the traffic model as follows:
\begin{enumerate}
\item {\em Diffusion \& Coalescence} - Consider a stationary car,
located at the left end of a free-flow domain. This car may
accelerate to the maximal velocity $v_{\max}$, reach the right
boundary and eventually stop there. This process corresponds to
the diffusion of the whole domain to the left (Fig.
\ref{fig.cartoon}a).
\item {\em Chipping} - A stationary car at the left end of a domain may
accelerate, and then brake before reaching the maximal velocity.
This process decreases the size of the domain, and ejects a single
vacancy to the left (Fig. \ref{fig.cartoon}b).
\end{enumerate}
Note that chipping occurs only to the left, 
corresponding to a fully asymmetric CM, with $\omega_R=0$.
\begin{figure}
\twofigures[width=6cm]{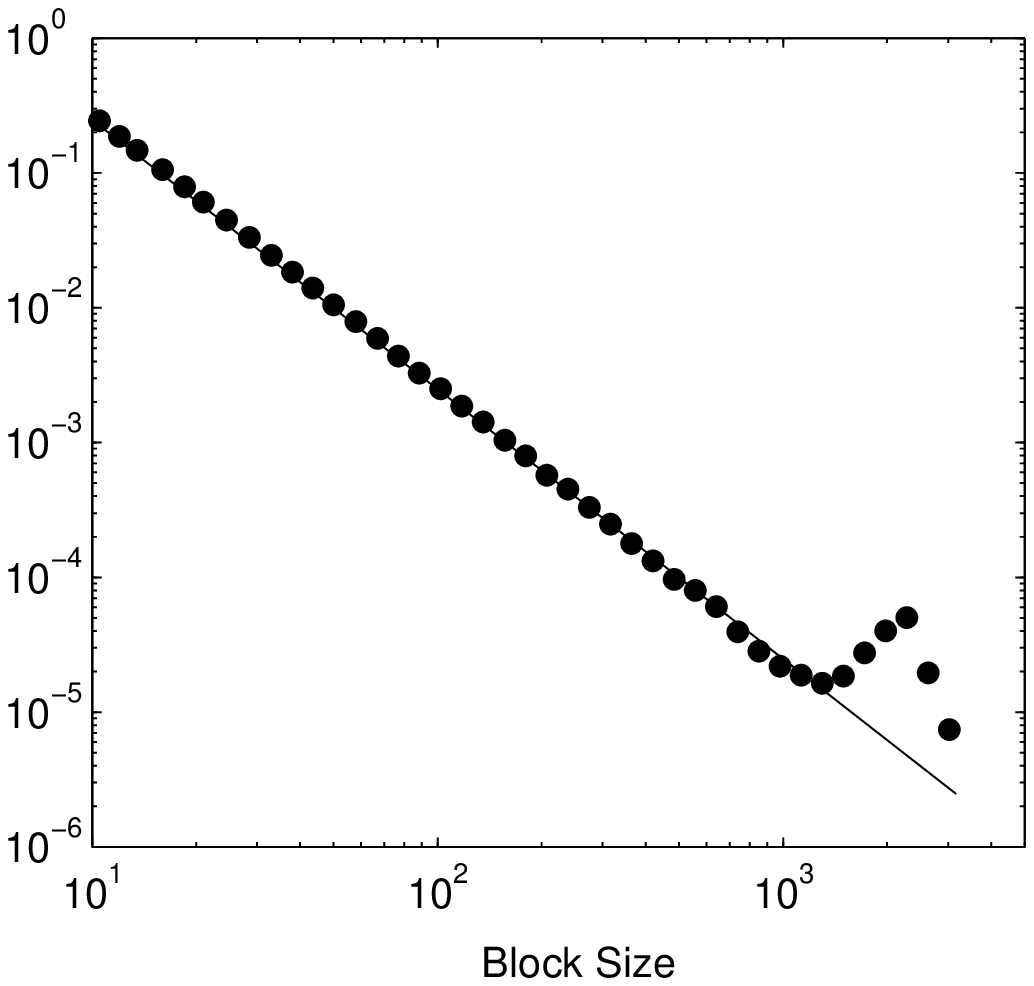}{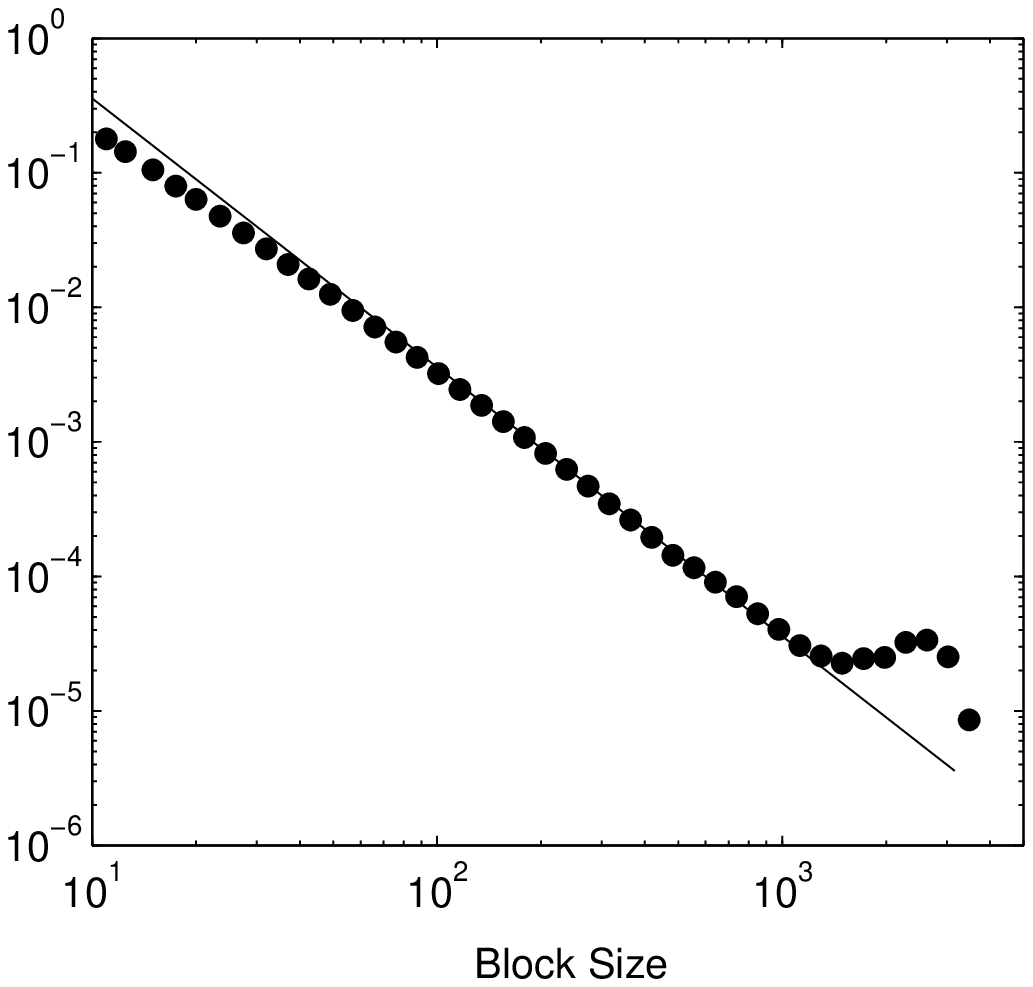} \caption{Domain
size distribution of VDB model with $v_{\max}=2,\rho=0.4,p=0.3$
and $q=0$. Solid line has a slope $-2$. Simulation was performed
on a system of size $L=10,000$ and averaged over $15 \times 10^7$
sweeps. \label{fig.3}} \caption{Domain size distribution of TCA
model with $\rho=0.4,\alpha=0.2,\beta=0.6,\gamma=0.8$ and
$\delta=1$. Solid line has a slope $-2$. Simulation was performed
on a system of size $L=10,000$ and averaged over $60 \times 10^6$
sweeps.\label{fig.4}}
\end{figure}

To test this picture we performed Monte-Carlo simulations of the
VDB model with  $q=0$ and $v_{\max}=2$, and measured the domain
size distribution. For convenience we define the size of a
free-flow domain, $k$, as the number of vacancies it
contains\footnote{This definition is not unique, as one may
include the deterministic cars residing in a domain in the
definition of the length, without changing the results. One
advantage of the definition used here is that the overall length
of domains is conserved.}. We find that the asymptotic form of
$P(k)$ is consistent with $k^{-\tau}$ with $\tau=2$
(Fig.~\ref{fig.3}), independent of the parameter $p$. In view of
the fact that $\tau = 2$, the existence of a macroscopic domain in
Fig.~\ref{fig.3}, as is demonstrated by the peak at large blocks,
should be interpreted as a finite-size effect, which would
disappear in the thermodynamic limit \cite{Rajesh02}.

It is interesting to note that unlike the CM, 
the number of domains in the traffic model is not conserved, but
is subject to fluctuations. The number of domains is given by
$M-M_D$ where $M_D$ is the number of deterministic
cars\footnote{Alternatively, one may associate each domain with a
stationary car, say, the one to its left. Within this definition
the number of domains is given by the number of stationary cars,
where domains of size zero are counted as well. Such a domain
corresponds to two adjacent stationary cars.
For the purpose of the present study each
definition can be used.}, \ie cars moving at $v=v_{\max}$. The
fluctuations in $M_D$ are expected to scale as $\sqrt{L}$. This
was verified by direct numerical studies of systems of size up to
$10^4$. Therefore, these fluctuations are not expected to
influence the dynamics of the system in the thermodynamic limit.

We conclude that in the CC limit of the VDB model there is no
phase transition for $\rho>\rho_0$. It is interesting to examine
the implication of this result on the question of existence of a
phase transition beyond the CC limit, namely for $q>0$. Here the
free-flow phase is no longer an absorbing state, and thus the
transition taking place at $\rho_0$ in the CC limit is expected to
become a smooth crossover. Since no other transition is found at
$\rho > \rho_0$ in the CC limit we also expect no such transition
for $q>0$. The reason is that in addition to the CM dynamics, a
model with $q>0$ exhibits other processes, which clearly disfavor
condensation of macroscopic domains. For example, here a domain
may split into fractions of comparable size, leading to fast
fragmentation of large domains.

In examining other traffic models, we find that in many cases the
dynamical processes characterizing the CM may still be used to
describe the domain dynamics. However, it may turn out that the
chipping process involves a detachment of more than a single
particle. Nevertheless, as long as the number of chipped particles
$r$ is bounded by a finite number, or the probability of chipping
$r$ particles $u(r)$ decays sufficiently fast with $r$ (say
exponentially), the main results obtained from the CM are expected
to be valid. Namely, condensation transition should not take place
as long as the chipping process is asymmetric. To verify this
point we studied numerically chipping models for the case $u(r)
\sim \exp(-r)$ and for the case where $r$ is distributed uniformly
over a finite range. In both cases we found that the domain size
distribution decays as $z^k/k^2$ for large $k$, as expected from
the CM. The VDB model with $v_{\max}>2$ corresponds to CM with
chipping of more than a single particle. However, in the CC limit
the number of chipped particles is bound by
$v_{\max}(v_{\max}-1)/2$. Thus the analysis presented above for
$v_{\max}=2$ remains valid for any finite $v_{\max}$.

To demonstrate the more general applicability of the CM picture to
traffic models, we briefly consider three other traffic models
which have been studied in the past.

{\em Model for Emergent Traffic Jams} \cite{Nagel95} --- This is
another variant of the Nagel-Screckenberg traffic model. The model
is closely related to the CC limit of the VDB model defined above,
except that the velocity update rule (Eq. \ref{eq.modela}) is
replaced by
\begin{eqnarray}
\label{eq.np}
v_i(t+\half)&=&\min\left\{v_i(t)+1,v_{\max},x_{i+1}(t)-x_i(t)-1\right\}
\nonumber \\
v_i(t+1)&=&\begin{cases}v_i(t+\half) &
\text{with probability } 1-p(v_i(t)) \\
\max\left\{0,v_i(t+\half)-1\right\} & \text{with probability }
p(v_i(t))
\end{cases}\,.
\end{eqnarray}
Unlike the VDB model, here non-deterministic cars slow down rather
then brake, and thus with $v_{\max}=1$ the two models are
identical. This model was shown to exhibit self-organized
criticality. Based on general considerations and numerical
simulations \cite{Nagel95}, it was argued that the size
distribution of the domains (termed holes in \cite{Nagel95})
behaves asymptotically as $k^{-2}$. Indeed, in \cite{Nagel95} the
evolution of domains was described in terms similar to those of
the CM. Note that in this case the chipping size distribution
$u(r)$ decays exponentially with $r$.

{\em Traffic Cellular Automata} \cite{Gray01} --- This model
belongs to a different class of CA traffic models where no
velocity variable is attached to a car. Cars move to their nearest
neighbor site with a probability that depends on the configuration
of their neighborhood. In \cite{Gray01} the dynamics is defined as
\begin{align}
\bullet\bullet\circ\circ&\mathop{\longrightarrow}^\alpha\bullet\circ\bullet\circ
&
\circ\bullet\circ\bullet&\mathop{\longrightarrow}^\beta\circ\circ\bullet\bullet
&
\bullet\bullet\circ\bullet&\mathop{\longrightarrow}^\gamma\bullet\circ\bullet\bullet
&
\circ\bullet\circ\circ&\mathop{\longrightarrow}^\delta\circ\circ\bullet\circ
\,,
\end{align}
where $\bullet$ denotes a car and $\circ$ a vacancy. In the
symmetric CC case, $\gamma=\delta=1$, the model exhibits a
low-density absorbing state at $\rho<1/3$ and a high-density
absorbing state at $\rho>2/3$. It has been suggested \cite{Gray01}
that for intermediate densities ($1/3 < \rho < 2/3$) and in some
region of the $\alpha,\beta$-plane, the system exhibits a
macroscopic jam, suggesting a jamming phase transition at some
density. The correspondence between this case and the CM is less
transparent, and will be addressed in a future publication
\cite{Ziv04}. Here we consider the CC limit, $\delta=1$, with
$\gamma<1$, and apply the CM approach to analyze the jammed phase.
In this phase a typical microscopic configuration is given by an
alternating left-to-right sequence of (a) free-flow regions,
composed of vacancies and cars separated from their nearest
neighbor cars by at least two vacancies, (b) finite {\em mixed}
region of alternating cars and vacancies, and (c) an uninterrupted
sequence of cars. In order to apply the approach described above,
it is convenient to define a domain as a union of adjacent
free-flow and mixed regions (a and b above). By examining the
dynamics of such domains one finds that they indeed exhibit the
characteristic processes of the CM \cite{Ziv04}. We performed
Monte-Carlo simulations of this model in the CC limit and measured
the domain size distribution (Fig.~\ref{fig.4}). We find that the
distribution is consistent with $k^{-2}$, as expected from the CM
picture, indicating again that a phase transition does not take
place in the jammed state.

It is interesting to note that if instead of $\delta=1$ we
consider $\gamma=1$, the role played by cars and vacancies is
interchanged. Here one can define a domain as a stretch of cars,
within which deterministic vacancies are embedded, followed to the
left by a mixed region. In this case the distribution of these
domains (or jams) behaves as $k^{-2}$, and thus no macroscopic jam
is expected.

{\em Traffic Model with Passing} \cite{Ispolatov95} --- Unlike all
models mentioned above, this model is not a cellular automaton.
The model is defined on a continuous ring, and evolves in
continuous time. Each car is assigned a-priory a random velocity
with which it moves on the ring. When a car encounters a slower
car it assumes its velocity, thus creating a jam. With some finite
probability the next to leading car in a jam can bypass its
predecessor and recover its original velocity. The direct
correspondence with the CM was already noted in
\cite{Ispolatov95}. Treating the CM within mean-field
approximation, the authors concluded that the model should exhibit
macroscopic jam. However, numerical simulations show that the jam
size distributions is again $k^{-2}$ \cite{Ispolatov95}. Our
approach suggests that in the observed distribution is related to
the asymmetric nature of the chipping, a feature which cannot be
captured in mean-field.
%

In summary, it is suggested that in many traffic models the
coarse-grained dynamics of domains (of either high or low density)
in some deterministic limit, may be described by the two basic
processes of the chipping-model, namely diffusion and chipping.
Analysis of several traffic models within this approach indicates
that as in the asymmetric CM, the traffic models do not exhibit a
jamming transition beyond perhaps the one related to the existence
of an absorbing state. It is concluded that in non-deterministic
traffic models jamming phase transitions do not take place.
Rather, a smooth crossover between a free-flow and a jammed state
takes place as the car density is increased. The approach outlined
in this paper could provide a useful tool for analyzing the
behavior of traffic models. In studying a specific model one first
has to establish (using numerical or other methods) that indeed
the coarse-grained dynamics of the domains does follow the basic
processes of the chipping model. Only then one can apply the
correspondence between the two. It would be of interest to test
the applicability of this approach to broader classes of traffic
models.

\acknowledgments We thank M.R. Evans and J.L. Lebowitz for useful
discussions. The support of the Israeli Science Foundation is
gratefully acknowledged.


\end{document}